\documentclass[12pt,preprint]{aastex}

\newcommand{\SgrA}{Sgr\,A$^\star$\,}
\newcommand{\rg}{ \ensuremath{r_{\rm g}} }
\newcommand{\rms}{ \ensuremath{r_{\rm ms}} }
\newcommand{\ui}{ \ensuremath{u_{\rm i}} }
\newcommand{\Ti}{ \ensuremath{T_{\rm i}} }
\newcommand{\mdot}{\ensuremath{\dot{M}_{\rm a}}}
\newcommand{\mdotw}{\ensuremath{\dot{M}_{\rm w}}}

\newcommand{\LEdd}{\ensuremath{L_{\rm Edd}}}
\newcommand{\Ld}{\ensuremath{L_{\rm d}}}
\newcommand{\ed}{\ensuremath{\epsilon_{\rm d}}}
\newcommand{\Pj}{\ensuremath{P_{\rm j}}}
\newcommand{\ej}{\ensuremath{\epsilon_{\rm j}}}
\newcommand{\Pa}{\ensuremath{P_{\rm a}}}
\newcommand{\ea}{\ensuremath{\epsilon_{\rm a}}}

\newcommand{\ent}{\ensuremath{\epsilon^{\scriptscriptstyle{\rm NT}}}}

\newcommand{\fnt}{\ensuremath{f^{\scriptscriptstyle{\rm NT}}(x)}}
\newcommand{\frphi}{\ensuremath{f^{\rm nzt}_{r \phi}(x)}}
\newcommand{\fphiz}{\ensuremath{f^{\rm nzt}_{\phi z}(x)}}

\newcommand{\xms}{\ensuremath{x_{\rm ms}}}
\newcommand{\yms}{\ensuremath{y_{\rm ms}}}
\newcommand{\nuobs}{\ensuremath{\nu_{\rm obs}}}
\newcommand{\sigmaT}{\sigma_{\scriptscriptstyle \rm T}}
\newcommand{\ltapprox}{\raisebox{-0.5ex}{$\,\stackrel{<}{\scriptstyle
\sim}\,$}}
\newcommand{\gtapprox}{\raisebox{-0.5ex}{$\,\stackrel{>}{\scriptstyle
\sim}\,$}}

\shorttitle{Jet-Driven Accretion in Sgr A$^\star$}
\shortauthors{Jolley \& Kuncic}

\begin{document}


\title{Constraints on jet-driven disk accretion in Sagittarius
  A$^\star$}


\author{Erin J. D. Jolley and Zdenka Kuncic}
\affil{School of Physics, University of Sydney,
   Sydney NSW, Australia}


\begin{abstract}
We revisit theoretical and observational constraints on geometrically-thin disk accretion in
Sagittarius\,A$^\star$ (\SgrA). We show that the combined effects of
mass outflows and electron energization in the hot part
of the accretion flow can deflate the inflowing gas from a
geometrically-thick structure. This allows the gas
to cool and even thermalize on an inflow timescale. As a result, a compact,
relatively cool disk may form at small radii. We show that magnetic coupling 
between the relativistic disk and a steady-state jet results in a 
disk that is less luminous than a standard relativistic disk accreting at the same rate. 
This relaxes the observational constraints on
thin-disk accretion in \SgrA (and by implication, other Low-Luminosity Active Galactic
Nulcei, LLAGN).
We find typical cold gas accretion rates of 
$\mbox{a few} \times 10^{-9} M_\odot \, {\rm yr}^{-1}$. 
We also find that the predicted modified disk emission is
compatible with existing near-infrared (NIR) observations of \SgrA in
its quiescent state provided that the disk
inclination angle is $\gtapprox 87^\circ$ and that the jet extracts
more than $75$\% of the accretion power. 
\end{abstract}





\section{Introduction}

The exceptionally compact radio source  \SgrA  is
spatially coincident with the dynamical centre of the Galaxy, which
contains a mass $M \approx 3.7 \times 10^6 M_\odot$, deduced from
measurements of stellar orbital motions
\citep[e.g.][]{b2,Schodel02,b1}. High-resolution VLBA imaging of \SgrA
at millimetre (mm)  wavelengths provides the most compelling observational
evidence yet for the existence of supermassive black holes (SMBHs), with $M$
now constrained to lie within a radius $\approx 1$\,AU of \SgrA \citep{b23}. 

The observed bolometric luminosity of \SgrA, $L \approx 10^{36}\,{\rm erg \,
s}^{-1} \approx 2 \times 10^{-9} \LEdd$ (where $\LEdd$ is the
Eddington luminosity), is unusually low, even compared to that
of other LLAGN \citep{b3}.
This can be attributed to a very low mass accretion rate, $\mdot$, or
a low radiative efficiency, $\epsilon = L/\mdot c^2$, or to a
combination of both. The most popular accretion models for \SgrA
are Radiatively Inefficient Accretion Flows (RIAFs), which have
attributed the low $L$ to a low $\epsilon$ and which are based on the
hot ion torus \citep*{SLE76,Ichimaru77,Rees82} and Advection Dominated
Accretion Flow (ADAF; \citealt{NarYi94}) models.

Applications of RIAFs to \SgrA have evolved considerably in
recent years in response to increasing observational constraints (see
\citealt{Yuan06} for a review). In the original models, the inferred
mass accretion rate was comparable to the Bondi rate for spherical
accretion,
$\simeq 10^{-5} \, M_\odot \, {\rm yr}^{-1}$, and the radiative efficiency was very
low, $\epsilon \simeq \mbox{a few} \times 10^{-6}$, as a result of
preferential ion viscous heating and negligible electron-ion coupling.
The most recent RIAF model \citep*{YQN03,YQN04} now takes into
consideration a mass outflow component as
well as energization of the electrons. As a result, the revised RIAF
model now attributes the
low luminosity of \SgrA to a combination of low \mdot \, and moderately
low $\epsilon$. The outflow reduces the rate at which hot ionized gas
is fed onto the central SMBH to levels that are consistent with mm and
sub-mm polarization measurements, which require
 $\mdot \ltapprox 4 \times 10^{-8} \, M_\odot \, {\rm yr}^{-1}$
\citep*{b16,b17,b18,Bower03,Marrone06,b26}. Electron heating and
acceleration are then required to produce the observed levels of
radio and sub-mm emission \citep{QuatNar99}, resulting in a radiative
efficiency $\epsilon \simeq 10^{-2}$.


The chanelling of binding energy away from the ions as a result of
outflows and electron heating (as well as electron-ion coupling)
necessarily implies that the ion energy is subvirial. There may then be
insufficient ion pressure to support a geometrically-thick structure and
consequently, some of the gas can radiatively cool on an inflow time
\citep{Rees82}. This raises the possibility that a compact, cool
accretion disk may form at small radii. To test this possibility, the
effects of a mass outflow on the continuity, angular momentum and
energy equations need to be self-consistently considered in the
exisiting RIAF model for \SgrA  \citep{YQN03}. Numerical
approaches \citep[e.g.][]{HawBalb02,SQHS07} have so far been limited by the
nonconservative numerical scheme and the neglect of radiative cooling.

Observationally, increasing evidence is emerging to suggest that geometrically-thin
disk accretion may persist at or near the last marginally stable
orbit in low-luminosity sources \citep{Miller06a,Miller06b,Maoz07,Rykoff07,Liu07}.
Young massive stars seen orbiting the Galactic Center (GC) and
believed to have formed \textit{in situ}
\citep{NaySun05,Nay06,Pau06,Nayakshin06,Lev06,Bel06} 
suggest that at least some of the hot gas detected by \textit{Chandra}
has condensed into a cool phase.
Additionally, new models for stellar wind
accretion in \SgrA indicate that slow winds can 
radiatively cool within a dynamical timescale to
produce a cold phase of accreting gas \citep{Cuadra06}. However, a strong constraint
on the presence of a cold, optically-thick disk is the absence of eclipses in the orbit of
S2, the best studied star in the central S-cluster. This limits the
size of a putative accretion
disk to $\ltapprox 10^{16} \, {\rm cm} \simeq 3 \times 10^5 \rg$ \citep*{Cuadra03}, where
$\rg = GM/c^2 \approx 5 \times 10^{11} (M/3.7\times 10^6 M_\odot) \, {\rm cm}$ is the
gravitational radius. 

If a compact, cool disk forms near \SgrA, much of its emission would
suffer strong interstellar extinction. Even so, the mass accretion
rate would have to be extremely low and the spin axes of the disk and
galactic plane would have to be closely aligned for the emission from
a standard disk to remain below observational limits in the NIR
\citep{Cuadra03,Trippe07}. However, it has been suggested
\citep*[e.g.][]{NagWilFal01,Falcke01,FKM04,Gallo07,MH07} that the
spectral properties of LLAGN and galactic X-ray binaries in their
low/hard X-ray state can be attributed to a jet-dominated mode of
accretion \citep[see also][]{MerlFab02}. Although jets have not been
observed in \SgrA, various arguments have been made for their
presence \citep*[see e.g.][for a comprehensive discussion]{MarkBowFalck07}.
If jets are present, an accretion disk can no longer be described by
the standard model \citep{b9,b10} because the radial disk structure is
modified by the magnetic torque responsible for jet formation
\citep{KB04,KB07b}. Indeed, extraction of accretion power by a magnetized jet results in an accretion disk that is less luminous than a standard disk accreting at the same rate \citep{KB07a}. That is, the radiative efficiency is lower than that predicted by standard accretion disk theory.

In this paper, we revisit theoretical constraints on the formation of a cool accretion disk  and we investigate observational constraints on jet-driven disk accretion in \SgrA. Unlike the jet-RIAF model for \SgrA \citep*{FalckMark00,YMF02} and the disk-corona model for LLAGN \citep{MerlFab02}, we explicitly model the magnetic coupling between the accretion flow and the outflow. This model has been successfully applied to M87 \citep{JK07}.
The organization of this paper is as follows.
In Section \ref{sec2}, we show that a RIAF becomes geometrically-thin
as a result of diversion of binding energy from the inflowing
ions. This allows the electrons to radiatively cool and to thermalize
with the ions on an inflow time, thus
forming a cool disk at small radii. We also present the relevant
equations for relativistic disk 
accretion modified by MHD stresses and we
calculate the modified disk flux radial profile and corresponding modified disk emission spectrum 
using parameters appropriate for \SgrA. In Section \ref{sec3}, we
calculate the steady-state synchrotron spectrum resulting
from a jet magnetically coupled to the underlying accretion flow. We
compare the predicted spectra with the observed quiescent
spectrum for \SgrA. A discussion and
concluding remarks are given in Sections \ref{sec4}  and \ref{Conclusions}, respectively. 

\section{Coupling a Magnetized Jet to a Relativistic Disk}\label{sec2}

\subsection{Formation of a Cool Disk}


Weakly magnetized, differentially rotating accretion flows are
unstable to MHD turbulence and electron
heating is unavoidable as a
result of viscous dissipation via a turbulent cascade as well as
resistive dissipation via stochastic reconnection
\citep{QuatGruz99,BKL00,SanInut01,SQHS07}. As the accretion flow
becomes increasingly radiatively efficient, the internal energy of the
ions drops to subvirial levels and
the accretion flow geometry deflates from a geometrically-thick
torus \citep{Rees82}. If an outflow is also present, then this
can further promote the collapse to a geometrically-thin structure, with a
height-to-radius ratio $h/r \ll 1$. 

To see this quantitatively, consider the internal energy equation for
ions, with specific energy $\ui$, in an axisymmetric, steady and
incompressible accretion flow with
radial velocity $v_r$ in
which a fraction $\delta$ of the binding energy extracted by the
internal MHD stresses $t_{r\phi}$ is diverted to the electrons and in
which there is mass outflow with velocity $v_z$:
\begin{equation}
\frac{1}{r} \frac{\partial}{\partial r} \left( r \rho \ui v_r \right)
+ \frac{\partial}{\partial z} \left( \rho \ui v_z \right)
= \frac{1}{2}  t_{r\phi} \, r \frac{\partial \Omega}{\partial r}
\end{equation} 
Note that the current RIAF model \citep{YQN03} neglects the outflow
term involving $v_z$. Vertically integrating and using the relations
$\mdot = 2\pi r \Sigma (-v_r)$ and
$\partial \mdotw /\partial r = -4\pi r \rho (h) v_z (h)$,
where $\mdotw$ is the mass loss rate and where continuity
implies $\partial \mdot /\partial r = -\partial \mdotw /\partial r$,
gives
\begin{equation}
\frac{d}{dr} \left( \mdot \ui \right) + \ui \frac{d\mdotw}{dr}
= -2\pi r T_{r\phi} \, r \frac{\partial \Omega}{\partial r}
\label{e:ui}
\end{equation}
where $T_{r\phi}$ is the vertically-averaged stress. This is obtained
from conservation of angular momentum:
\begin{equation}
-2\pi r T_{r\phi} = \mdot r \Omega 
\left[ 1 - \frac{\mdot (\rms)}{\mdot (r)} \left( \frac{r}{\rms}
  \right)^{-1/2} \right]
- \frac{1}{r} \int_{\rms}^r r^2 \Omega \frac{d\mdot}{dr} \, dr
\end{equation}
where $\rms$ is the last marginally stable orbit of the inflow. Substituting
this into (\ref{e:ui}) to eliminate $T_{r\phi}$ and using $\mdot
\propto r^s$ \citep{YQN03}, where $0 < s < 1$ and a keplerian angular
velocity, $\Omega = (GM/r^3)^{1/2}$, yields the following solution for
the ion internal energy:
\begin{equation}
u_i (r) = \frac{3}{2} \left[ \frac{\frac{1}{2} (1-\delta)}{s +
    \frac{1}{2}} \right] \zeta (r) \frac{GM}{r}
\end{equation}
where $\zeta (r) = 1 - (3/2 + s)^{-1} (r/\rms)^{-1/2+s}$ is a small-$r$
correction factor. This solution can be written in terms of the ion
temperature $\Ti$ and the ion virial temperature $T_{\rm vir,i}$:
\begin{equation}
\frac{\Ti}{T_{\rm vir,i}} \approx \frac{\frac{1}{2}}{s + \frac{1}{2}}
\left( 1 - \delta \right)
\end{equation}
Thus, the ions are subvirial. The current RIAF model for \SgrA
requires $\delta \gtapprox 0.5$ and numerical simulations
\citep[e.g.][]{HawBalb02} indicate $s
\approx 0.5 - 1$. Taking $s=0.75$ for example, gives $\Ti/T_{\rm vir,i} \ltapprox
0.2$.

This result has important implications for the radiative cooling and
electron-ion collision rates, both of which are very sensitive to the
scaleheight ratio $h/r \approx \Ti/T_{\rm vir,i}$ of an
ion-pressure-supported disk.
Consider the inflow timescale, $t^{\rm inflow} \simeq r/|v_r|$, and
the bremsstrahlung cooling timescale
\citep[e.g.][]{RL},
\begin{equation}
t^{\rm cool} \simeq 7 \times 10^{10} n_{\rm e}^{-1} T_{\rm e}^{1/2} \> {\rm s} \qquad 
\label{e:tcool}
\end{equation}
where $n_{\rm e}$ is the electron number density (in units ${\rm
  cm}^{-3} $) and $T_{\rm e}$ is the electron temperature (in $\rm
K$). Using $n_{\rm e} \approx \mdot /(4\pi r
\mu m_{\rm p} h |v_r|)$, the ratio of cooling to inflow timescales is
\begin{equation}
\frac{t^{\rm cool}}{t^{\rm inflow}} \simeq 100 \, \frac{\epsilon}{\dot m}
\frac{h}{\rg} \frac{v_r^2}{c^2}
\left( \frac{kT_{\rm e}}{m_{\rm e}c^2} \right)^{1/2}
\end{equation}
where 
$\dot m \equiv
\mdot / \dot M_{\rm Edd}$ is the dimensionless mass accretion
rate and $\dot M_{\rm Edd} = L_{\rm
  Edd}/(\epsilon c^2)$ is the Eddington accretion rate\footnote{Note
  that we define $\dot M_{\rm Edd}$ without assuming a 10\% radiative
  efficiency, as is done in RIAF models, in order to keep the
  dependence on $\epsilon$ explicit throughout the equations.}.
Using the relation $|v_r| = \frac{3}{2} \alpha c_{\rm s}
\frac{h}{r}$ from the $\alpha$-disk formalism, where $c_{\rm s} \approx
(\gamma k\Ti/\mu m_{\rm p})^{1/2} = \Omega h$ is the isothermal sound
speed, the condition $t^{\rm cool} / t^{\rm inflow} \ltapprox 1$
required for the electrons to cool
before reaching the black hole then implies \citep[see also e.g.][]{Rees82}
\begin{equation}
\dot m \gtapprox 2 \times 10^{-9} \, \epsilon_{-2} \, \alpha_{-2}^2
\left( \frac{h}{0.1 r} \right)^5
\left( \frac{kT_{\rm e}}{m_{\rm e}c^2} \right)^{1/2}
 \qquad 
\end{equation}
where $\epsilon_{-2} = \epsilon/10^{-2}$ and $\alpha_{-2} = \alpha
/10^{-2}$. Typical values of $\alpha$ found in numerical simulations
are a few percent \citep[see][]{Balbus03}. Note that $\dot m =  L/\LEdd
\approx 2 \times 10^{-9}$ for \SgrA, implying that hot electrons with
initial temperatures $T_{\rm e} \ltapprox 10^{10}\,$K can cool down by the
time they reach the black hole.

If radiative cooling by bremsstrahlung emission resulting from
electron-ion encounters can occur on an inflow time when the ions are
subvirial, then it is necessary to also reconsider whether the
Coulomb collision time can be shorter than the inflow time. The
electron-ion collision time, $t^{\rm ei} \approx (\pi/2)^{1/2} m_{\rm
  p}/(m_{\rm e} n_{\rm e} \sigmaT c \ln \Lambda ) (kT_{\rm e}/m_{\rm
  e} c^2 )^{3/2}$, can be written in terms of accretion parameters as
\begin{equation}
t^{\rm ei} \approx \left( \frac{\pi}{2} \right)^{1/2}
\frac{1}{\ln \Lambda} \frac{\epsilon}{\dot m} \frac{h |v_r|}{c^2}
\frac{m_{\rm p}}{m_{\rm e}} \frac{r}{\rg}
\left( \frac{kT_{\rm e}}{m_{\rm e}c^2} \right)^{3/2}
\end{equation}
The condition $t^{\rm ei} \ltapprox t^{\rm inflow}$ then requires
\begin{equation}
\dot m \gtapprox 50 \, \epsilon \, \alpha^2 \left( \frac{h}{r} \right)^5
\left( \frac{kT_{\rm e}}{m_{\rm e}c^2} \right)^{3/2}
= 5 \times 10^{-10} \, \epsilon_{-2} \, \alpha_{-2}^2
\left( \frac{h}{0.1 r} \right)^5
\left( \frac{kT_{\rm e}}{m_{\rm e}c^2} \right)^{3/2}
\end{equation}
where a Coulomb logarithm $\ln \Lambda = 50$ has been used.
Note that this differs from previous calculations
\citep[e.g.][]{Rees82} only in the explicit inclusion of $h/r$
(previously taken as unity) and $\epsilon$ (previously taken as
$0.1$). For \SgrA, with $\dot m \approx 2 \times 10^{-9}$, this
implies electrons with temperatures $T_{\rm e} \ltapprox 2 \times
10^{10}\,$K can thermally couple with subvirial ions on an inflow time
via Coulomb collisions. Note that collisionless
wave-particle plasma microinstabilities can enhance the electron-ion
coupling rate further
\citep{BegChi88,Quat98,Gruzinov98,QuatGruz99,Blackman99,BKL00,Melia01}.

As the disk cools, its scaleheight continues to decrease
until the gas becomes optically thick. Initially, the opacity
of the hot gas is dominated by electron scattering. The
optical depth over the disk height is $\tau_{\rm es} \approx \sigma_{\rm T} \mdot/(4\pi r \mu m_{\rm p} |v_r|)$, which exceeds unity when
\begin{equation}
\mdot \gtapprox 7 \times 10^{-7} \alpha_{-2} \left( \frac{r}{100 \rg}
\right)^{1/2} \left( \frac{h}{0.1 r} \right)^2 M_6 \> M_\odot \,
      {\rm yr}^{-1} \qquad 
\end{equation}
For \SgrA, this implies a disk can become optically-thick once it
collapses to scaleheights
$h \ltapprox \mbox{a few} \, \times 10^{-3} r$. By then, however, the
disk is sufficiently cool and dense that other opacities are more
important than electron scattering, so the final disk height need not
be too small.

We note that whilst numerical simulations
\citep[e.g.][]{SanInut01,SQHS07} demonstrate that
preferential heating of electrons resulting from viscous and resistive
turbulent dissipation occurs in low-$\mdot$ accretion flows, they are yet to
include radiative cooling in the calculations. This is needed in order to
verify the above theoretical arguments. In the following, we assume
that a cool, quasi-thermal disk has formed at small radii as a result
of electron heating and mass outflows in the hot accretion flow at
larger radii. A thermally-driven outflow is assumed to be negligible in
the cool disk, but we consider the effects of a magnetically-driven
outflow which we identify as a magnetized jet.

\subsection{Magnetic Torque on the Disk Surface}\label{NZTatsurface}
\label{sec:nzt}

The MHD stresses responsible for allowing accretion to proceed can give rise to a non-zero torque that acts over the disk surface, in addition to a non-zero torque at the last
marginally stable orbit \citep{KB04}. The effect of a non-zero magnetic
torque acting on the surface of a relativistic disk has not been
self-consistently modelled before. This torque does work against the disk
surface, thereby removing energy from the disk and directing it
vertically. Because it is associated with the azimuthal-vertical
magnetic stress $t_{\phi z}^+ = B_{\phi}^+B_z^+ / 4\pi$ at the disk
surface (denoted by the '+' superscript), it can be identified with
the formation of a magnetized corona and/or jet.
Thus, the radiative flux from a relativistic, torqued disk can be most
generally expressed as (see the Appendix for details)
\begin{equation}
F (r) = \frac{3GM\mdot}{8\pi r^3} \, \left[ \, \fnt + \frphi - \fphiz \, \right]
\label{fluxeq}
\end{equation}
where $x = r/\rg$ is the dimensionless radius, $\fnt$ is the \citet{b10} relativistic correction
factor, $\frphi$ is a correction factor for a nonzero torque (NZT) at
the inner disk boundary \citep{b12}, and $\fphiz$ is an analogous
correction factor for a nonzero torque on the disk surface. This is derived in
Appendix \ref{A2} and is explicitly given by
\begin{equation}
\fphiz = \frac{4\pi {\rm r_g}}{C(x) \mdot \Omega x^2}
\int^x_{\xms} \frac{C^{1/2}}{B} t^+_{\phi z} x^2 \, \mathrm{d}x \qquad
\end{equation}
where $\xms = r_{\rm ms}/\rg$ is the dimensionless radius of the last
marginally stable orbit and
the dimensionless functions $C(x)$ and $B(x)$ are defined in the
Appendix. 

Henceforth, we consider a jet interpretation for the nonzero surface
torque and assume a simple power law radial profile for the $\phi z$
stress as follows \citep[see also][]{Freeland06}:
\begin{equation}
t^+_{\phi z}(x) = t^+_{\phi z}(\xms) \left( \frac{x}{\xms} \right) ^{-q}
\label{tphiz}
\end{equation}
The normalization of the nonzero MHD stresses $t^+_{\phi z}(\xms)$ and
$T_{r \phi}(\xms)$ are yet to be determined. We now achieve this by
considering the global constraint imposed by energy conservation.

\subsection{The Global Energy Budget}

Global energy conservation requires that the total accretion power
$\Pa$ must equal the sum of the disk radiative power $\Ld$
and the jet power $\Pj$, which we equate to the rate at which
work is done against the disk by the nonzero magnetic torque on its
surface. Hence,
\begin{equation}
\label{e:global}
\Pa = \Ld + \Pj
\end{equation}
We now write
\begin{eqnarray}
\Pa &=& \ea \mdot c^2 = \left( \ent + \Delta \epsilon \right) \mdot c^2 \nonumber \\
\Pj &=& \ej \mdot c^2
\end{eqnarray}
where $\ea =  \ent + \Delta \epsilon$ is the total accretion efficiency 
and $\ej$ is the efficiency of energy removal to a jet. 
The efficiencies are defined as follows: 
\begin{equation} \label{epsnt}
\ent  = \frac{3}{2} \int_{\xms}^\infty  x^{-2} \fnt \,
 \mathrm{d}x
= \frac{3}{2} \int_{\xms}^\infty  x^{-2} 
  \frac{A(x)}{C(x)} \,
 \mathrm{d}x
\end{equation}
\begin{equation} \label{epsrphi}
\Delta \epsilon =  \frac{3}{2} \int_{\xms}^\infty  x^{-2} \frphi \,
\mathrm{d}x 
= \frac{C_{\rm ms}^{1/2}}{B_{\rm ms}} \frac{3\pi x_{\rm
    ms}^2 T_{r \phi}(\xms) \rg}{\mdot c} \, I_3 
\end{equation}
\begin{equation} \label{pjepsj}
\ej  = \frac{3}{2} \int_{\xms}^\infty x^{-2} \fphiz \, \mathrm{d}x
= \frac{6\pi \xms^q \rg^2 t_{\phi z}^+ (\xms)}{\mdot c} \, I_2
\end{equation}
where
\begin{equation}
I_1(x) = \int_{\xms}^x \frac{[C(x)]^{1/2}}{B(x)} x^{2-q} \, {\rm d}x
\end{equation}
\begin{equation}
I_2 = \int_{\xms}^{\infty} \frac{x^{-5/2}}{C(x)} \, I_1(x) \, {\rm d}x
\end{equation}
\begin{equation}
I_3 = \int_{\xms}^\infty \frac{x^{-5/2}}{C(x)} \, {\rm d}x
\end{equation}
Thus, the nonzero torques can be normalized in terms of the
corresponding efficiencies $\Delta \epsilon$ and $\ej$ and the terms
$\frphi$ and $\fphiz$ can be rewritten 
as follows:
\begin{equation}
\frphi = \frac{2}{3} \frac{\Delta \epsilon}{C(x) x^{1/2} I_3}
\end{equation}
\begin{equation}
\fphiz = \frac{2}{3} \frac{\ej}{C(x) x^{1/2}} \frac{I_1(x)}{I_2}
\end{equation}

\begin{figure}
\centerline{\includegraphics[width=8.5truecm]{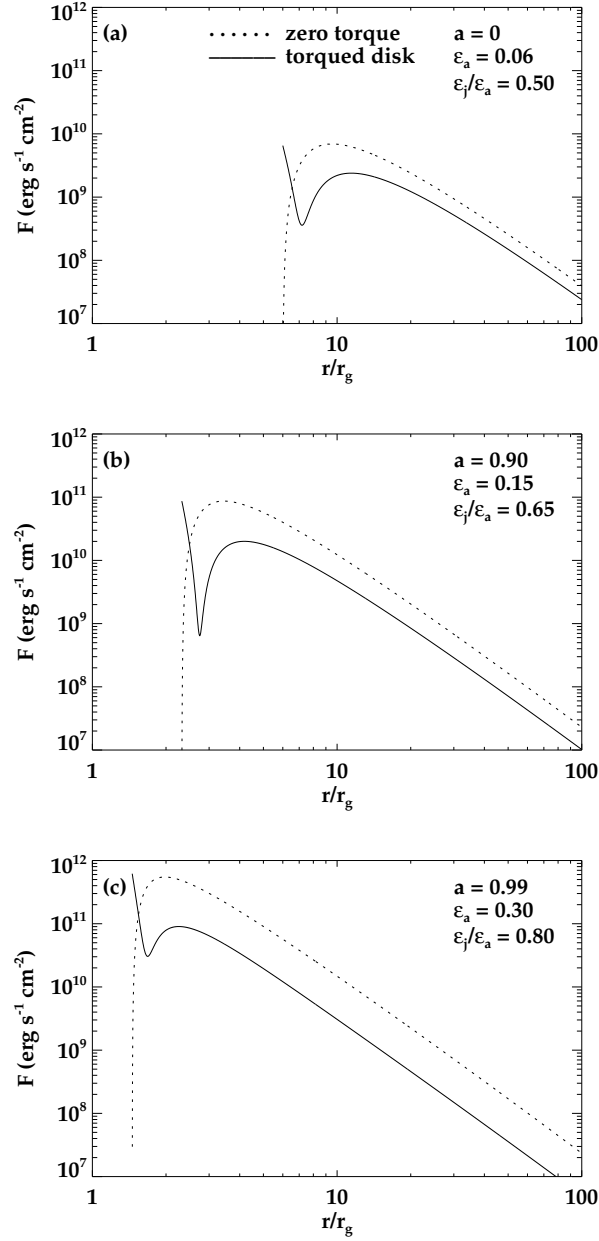}}
\caption{Radial flux profiles for the jet-modified disk model (solid
  curves) with
  different physical parameters:  $a$ is the dimensionless black hole
  spin parameter;  $\ea$ is the overall accretion efficiency, which
  includes a fractional contribution of $5$\% of the Novikov-Thorne
  efficiency resulting from a nonzero torque at the last marginally
  stable orbit $\rms$; and  $\ej/\ea$ is the  fractional power removed
  by a surface magnetic torque responsible for jet formation. The disk luminosity is $\Ld = 10^{-8} \LEdd \approx 5 \times 10^{36} {\rm erg \, s}^{-1}$ in all cases. The
  dotted curves show the corresponding radial flux
  profiles for a standard, non-torqued relativistic disk.}
\label{fluxfig}
\end{figure}

\subsection{The Modified Disk Flux Radial Profile and Spectrum}\label{last}

We now substitute the above expressions  into the disk flux profile (\ref{fluxeq}):
\begin{equation}\label{Fd}
 F (r) = \frac{3GM\mdot}{8\pi r^3}
\left[ \frac{A(x)}{C(x)} + \frac{2}{3} \frac{\Delta \epsilon}{C(x) x^{1/2} I_3}
-  \frac{2}{3} \frac{\ej}{C(x) x^{1/2}} \frac{I_1(x)}{I_2}  \right] 
\end{equation}
The global energy constraint (\ref{e:global}) yields the following for
the disk radiative efficiency:
\begin{equation}
\ed = \left( \ent + \Delta \epsilon \right) \left( 1 - \frac{\ej}{\ea} \right)
\end{equation}
We can also write
\begin{equation}
\mdot = \left( \frac{\Ld}{\LEdd} \right) \frac{\LEdd}{\ed c^2}
\label{e:mdot}
\end{equation}
where $\LEdd = 4\pi GM m_{\rm p} c / \sigmaT \approx 1.3 \times
10^{44} M_6 \, {\rm erg \, s}^{-1}$. Thus, once $\Delta
\epsilon$ and $\ej / \ea$ are specified (and note that $\ent$ is given
explicitly by (\ref{epsnt}) and depends only on the dimensionless spin
parameter $a$), we have determined $\ed$. Then the
mass accretion rate is determined by specifying $\Ld / \LEdd$ and using (\ref{e:mdot}). 
The fraction $\ej/\ea$
of accretion power injected into a jet cannot be
arbitrarily large, however, since the local disk flux must remain
positive at all radii. So for a given set of input parameters, $a$, $\Delta
\epsilon/\ent$,  and $\Ld / \LEdd$, there is a maximum allowable value
of $\ej / \ea$.

The radial flux profiles predicted by our torqued relativistic disk
model (c.f.~(\ref{Fd})) are shown in Figure~\ref{fluxfig} (solid curves). 
In all cases, the rate of radial decline in the surface magnetic
stresses is fixed at $q=2.5$ (c.f. 
(\ref{tphiz})). The efficiency of energy dissipation by the magnetic torque at the inner
boundary is set to $\Delta \epsilon = 0.05 \ent$. 
The intrinsic luminosity of the modified disk is $\Ld = 10^{-8} \LEdd \approx 5 \times 10^{36}\,{\rm erg
  \, s}^{-1}$ in all cases.
The flux profiles are calculated for a
range of black hole spin parameters $a$ and fractional jet powers $\ej
/ \ea$, as indicated in each plot. Note that as $a$, and therefore
$\ea$, increase, a larger fraction $\ej/ \ea$ of the accretion power
must be electromagnetically extracted to maintain a constant
$\Ld$. The mass accretion rates are as follows:
(a) $\mdot \approx 3 \times 10^{-9} M_\odot \, {\rm yr}^{-1}$,
(b) $\mdot \approx 1 \times 10^{-9} M_\odot \, {\rm yr}^{-1}$ and
(c) $\mdot \approx 1 \times 10^{-9} M_\odot \, {\rm yr}^{-1}$. 
The corresponding profiles for a non-torqued disk (dotted curves)
are calculated using the same values of $\mdot$, but in that case,
$\ed \approx \ea$ because all the accretion power is 
dissipated locally in the disk. 

The effect of the nonzero torque
across the disk surface is to do work against the disk, thus reducing
the disk flux over the range of radii where the magnetic torque is
strongest. This counteracts the effect of the nonzero torque at the
inner disk boundary, which enchances the disk flux near $\xms$. The
combined effects of these two torques is clearly evident in the radial
flux profiles in Fig.~\ref{fluxfig} (solid curves). The resulting 
radial profiles are substantially modified from their corresponding
zero-torque profiles (Fig.~\ref{fluxfig}, dotted lines). 
It is clear from Fig.~\ref{fluxfig} that the nonzero magnetic torque 
acting on the disk surface results in a disk radiative efficiency $\ed$ 
that is lower than that of a non-torqued disk. 

\begin{figure}
\centerline{\includegraphics[width = 8.5truecm]{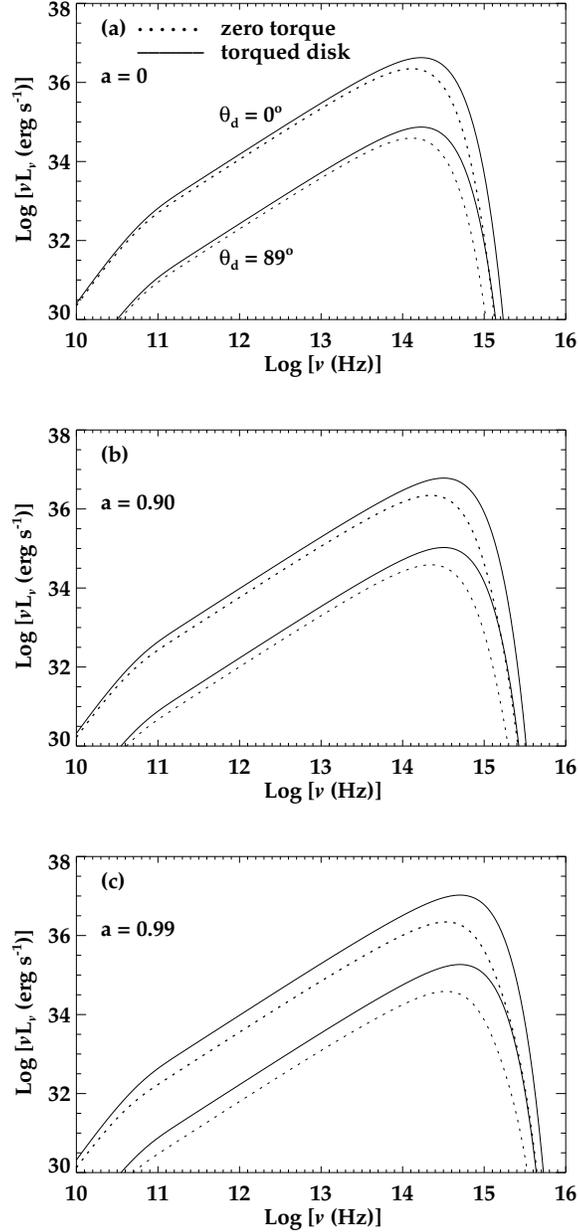}}
\caption{Predicted disk spectra for the jet-modified disk model. The
  dimensionless black hole spin parameter $a$ is and  $\theta_{\rm d}$
  is the inclination angle of the disk. All other model parameters are
  the same as those used in the corresponding flux profiles shown in
  Fig.~\ref{fluxfig}. The solid curves are the predicted spectra for a
  relativistic disk torqued at the inner boundary and on its surface;
  the dotted curves are spectra for a non-torqued relativistic disk
  (i.e. a Novikov-Thorne disk). In each plot, the upper two curves are for 
a disk oriented face-on ($\theta_{\rm d} = 0$) and the lower two
  curves are for a disk oriented nearly egde-on ($\theta_{\rm d} = 89^\circ$).}
\label{spectfig}
\end{figure}

Figure~\ref{spectfig} shows the disk spectra corresponding to
the flux profiles in  Fig.~\ref{fluxfig}. The spectra have been
calculated assuming local blackbody emission in each annulus of the
disk, with the local temperature determined by $T(r) =
[F(r)/\sigma]^{1/4}$, where $F (r)$ is given by (\ref{Fd}) and $\sigma$
is the Stefan-Boltzmann constant. We have included a $\cos \theta_{\rm
  d}$ projection factor, where $\theta_{\rm d}$ is the angle between
the disk spin axis and our line of sight.  Note that in all cases, the
effect of the MHD torque on the disk surface is to make the emergent spectrum
(solid curves) dimmer and redder than that of a non-torqued disk accreting at the
same rate (dotted curves). The effect is strongest for the high spin
case (Fig.~\ref{spectfig}c) because the overall accretion efficiency
is highest in that case and to maintain the same $\Ld$, the jet must remove a larger fraction of the
total accretion power.

\section{Jet Synchrotron Emission}\label{sec3}

We identify the nonzero magnetic torque across the disk surface with
the mechanism responsible for extracting accretion power from the disk
and injecting it directly into a jet. This torque could also
produce a magnetized coronal outflow, in which case only a fraction of
the extracted power may result in a jet \citep[see e.g.][]{MerlFab02}.
Some of the magnetic energy is subsequently converted into kinetic energy. 
We expect a fraction of the particles to be accelerated to nonthermal, relativistic energies. 
Synchrotron radiation by relativistic electrons will then contribute significantly to the 
radio emission. It has been proposed  \citep{FalckMark00} that highly efficient acceleration of electrons and synchrotron 
emission at the base of the jet may explain the observed sub-mm excess
emission from \SgrA.

\subsection{The Jet Model}\label{synchrotron}

\begin{figure}
\centering
\includegraphics[width=9truecm,clip=true,trim=350 200 30 0]{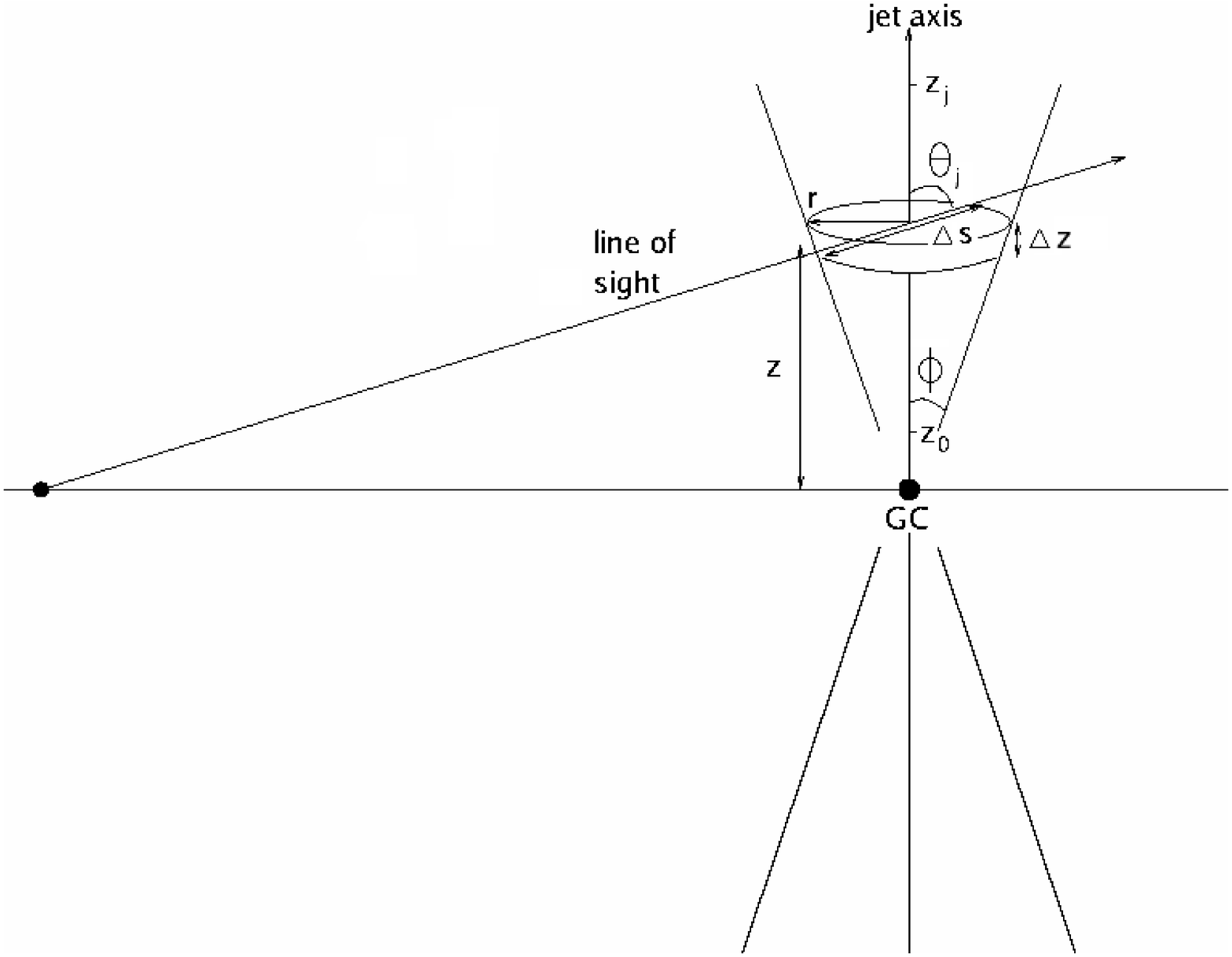}
\caption{Schematic diagram of the jet geometry (not to scale). The jet begins at a height $z_0$ and extends to $z_{\rm j}$, with 
half opening angle $\phi$. 
It is divided into cylindrical sections of thickness $\Delta z$ and radius $r \approx z \phi$. 
The angle between the jet axis and the line of sight is $\theta_{\rm j}$ and the path length 
through each cylinder is $\Delta s$. The distance to the galactic centre is ${\rm D_{GC}} \approx 8.5 \, {\rm kpc}$.}
\label{jetfig}
\end{figure}

We divide the jet into a series of quasi-cylindrical sections of thickness $\Delta z$, 
and calculate the total emission spectrum 
by summing the contributions from each component.
The geometry is illustrated in Figure~\ref{jetfig}.

The jet plasma is assumed to have a magnetic field strength $B$ and
to contain nonthermal electrons, with Lorentz factors 
in the range $\gamma_{\rm min} \le \gamma \le \gamma_{\rm max}$, where
$\gamma_{\rm min}$ and $ \gamma_{\rm max}$ are the minimum and maximum electron Lorentz
factors respectively.
The nonthermal electron energy distribution is given by 
$N_{\rm e}(\gamma) \propto \gamma^{-p}$ where $N_{\rm e} = \int N_{\rm e}(\gamma) \, {\rm d}\gamma$ 
is the total electron number density. 

We consider a relativistic jet with bulk Lorentz factor $\Gamma_{\rm j}$,  half-opening angle 
$\phi \approx 1/\Gamma_{\rm j}$, and Doppler factor 
\begin{equation}
\delta = \left\{ \Gamma_{\rm j} \left[1 - (1-\Gamma_{\rm
    j}^{-2})^{1/2} \cos \theta_{\rm j} \right] \right\}^{-1}
    \qquad ,
\end{equation}
where $\theta_{\rm j}$ is the angle between our line of sight and the jet axis (see Fig.~\ref{jetfig}).
We use the following simple radiative transfer model to calculate the observed specific 
luminosity due to the net contribution from each jet component 
(assuming isotropic emission in the source rest frame): 
\begin{equation} \label{L}
L_{\rm \nuobs}^{\rm obs} \approx 2\sum^{z_{\rm j}}_{z = z_0}  
4 \pi \delta^3 S_{\rm \nuobs}^{\rm syn}\left( 1- {\rm e}^{-\tau_{\rm \nuobs}^{\rm syn}} \right) \Delta A 
\end{equation}
where $\Delta A \approx \pi r \Delta z \sin \theta$ is the projected surface area of each emitting cylinder, 
$S_{\rm \nuobs}^{\rm syn}$ is the synchrotron source function (see e.g. \citealt{RL} for relevant formulas) and 
$$\tau_{\rm \nuobs}^{\rm syn} = \delta^{-1} \kappa_{\rm \nuobs}^{\rm syn} \Delta s$$ 
is the synchrotron optical depth along a path length 
$\Delta s \approx 2r / \cos\theta_{\rm j}$, where $r \approx z \phi$ through each section. 

We use an equipartition factor 
$
f_{\rm eq} = U_B/U_{\rm e} 
$
to relate the magnetic field energy density $U_B = B^2/8\pi$ 
directly to the relativistic electron
energy density $U_{\rm e} = \frac{4}{3} \langle\gamma\rangle N_{\rm e} m_{\rm e} {\rm c}^2$, where 
$\langle \gamma \rangle$ is the average
Lorentz factor. 
The electron number density $N_{\rm e}$ and hence the magnetic field $B$ 
decline with jet height $z$ according to 
$$N_{\rm e}(z) \propto z^{-2} \qquad {\rm ,} \qquad B(z) \propto z^{-1}$$
The total jet power is 
\begin{eqnarray}
\Pj && \approx \pi r_{\rm j}^2 \Gamma_{\rm j} (1 - \Gamma_{\rm j}^{-2})^{1/2} {\rm c} \times \nonumber \\
 && \left[ (\Gamma_{\rm j} - 1)N_{\rm e} m_p {\rm c}^2 + \frac{4}{3} \Gamma_{\rm j} N_{\rm e} \langle\gamma\rangle m_{\rm e} {\rm c}^2 \left( 1+ 2f_{\rm eq} \right) \right]
\end{eqnarray}
where the first term in square brackets refers to the bulk jet kinetic energy 
and the second term refers to the electron kinetic energy and the magnetic energy.
This is used to normalize the electron number density $N_{{\rm e},0 }$ at the base of the jet
once the ratio $\ej / \ea$ (or equivalently $P_{\rm j} / P_{\rm a}$) is specified. 

\subsection{Application to \SgrA}\label{saga}

\begin{figure*}
\centerline{\includegraphics[width = 16cm]{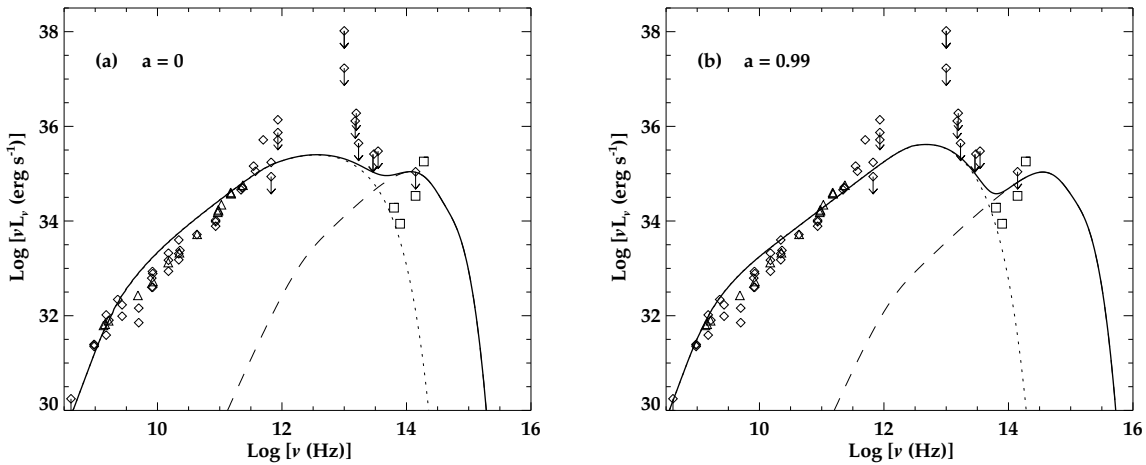}}
\caption{Broadband steady-state spectra predicted by the magnetically coupled
  disk-jet model for (a) a zero-spin black hole and (b) a high-spin
  black hole, compared against dereddened observations of \SgrA in
  quiescence. The solid line is the total disk+jet spectrum predicted
  by our model, the dotted line is the jet synchrotron spectrum and
  the dashed line is the modified disk spectrum for an inclination of
  $\theta_{\rm d} =87^\circ$.  The diamonds, triangles and squares are  data
  points compiled by \citet{b8}, \citet{b25} and \citet{BN06}, respectively.}
\label{datafig}
\end{figure*}

Figure~\ref{datafig} shows the best-fit broadband spectra predicted by our
magnetically coupled disk-jet model plotted against data points for
observations of \SgrA in quiescence \citep[compiled by][]{b8,b25,BN06}. The
best-fit disk parameters are listed in Table~1. The disk
inclination angle is $\theta_{\rm d} = 87^\circ$ in both the zero-spin
and high-spin cases. We generally find that the jet needs to
extract the maximum fraction $\ej / \ea$ of accretion power from the
disk to produce an energetically significant spectral component in the sub-mm. For
the jet spectrum, we used $\theta_{\rm j} = 55^\circ$
\citep{FalckMark00} and we found that a total jet length of just $z_{\rm j} \approx 350 \rg
\simeq 6 \times 10^{-5} \, {\rm pc}$ was needed to fit the lowest radio
frequency data points (this corresponds to an angular size of $\theta \simeq 1.5 \, {\rm mas}$ at the distance to \SgrA). The jet is launched from an initial height $z_0
= 0.1 \rms$ above the disk midplane, where $\rms = 6\rg$ for $a=0$ and $\rms = 1.5\rg$ for $a=0.99$. The electron energy power-law
index is $p=1.9$. The other jet input parameters used are as follows:
(a) ($a=0$ case) $\Gamma_{\rm j} =1.05$, $f_{\rm eq}=30$, $\gamma_{\rm
  min} = 30$, $\gamma_{\rm max}=150$; and
(b) ($a=0.99$ case) $\Gamma_{\rm j} =1.9$, $f_{\rm eq}=1$, $\gamma_{\rm
  min} = 50$, $\gamma_{\rm max}=100$. The corresponding electron number density and magnetic field strength at the base of the jet are: (a) $N_{{\rm e},0 } \approx  1 \times 10^6 \, {\rm cm}^{-3}$, $B_0 \approx 230 \, {\rm G}$; and (b) $N_{{\rm e},0 } \approx  1 \times 10^8 \, {\rm cm}^{-3}$, $B_0 \approx 430 \, {\rm G}$. Despite the large difference in $N_{{\rm e},0 }$, the predicted jet spectra for the two cases are remarkably similar. This is because the lower $N_{{\rm e},0 }$ for the $a=0$ case is offset by the larger length scale, since we set $z_0 = 0.1 \rms$ in both cases and $\rms$ is larger by a factor of $4$ for the $a=0$ case. The location of the synchrotron self-absorption turnover is sensitive to the value of $N_{{\rm e},0 }$ and we find that $z_0 = 0.1 \rms$ gives a turnover in the sub-mm in both cases, which is  suggested by the observations.
  
Note that in both the $a=0$ and $a=0.99$ cases, we find that a relatively narrow range of electron Lorentz factors, $\gamma_{\rm min}\ltapprox \gamma \ltapprox \gamma_{\rm
  max}$, is required to produce a prominent ``bump'' feature in the sub-mm.
The range of electron energies is not sufficiently wide to produce a
broadband optically-thin synchrotron power-law spectrum, so the spectrum
cuts off sharply above the maximum critical frequency at the base of the jet,
$\sim 4 \times 10^6 \gamma_{\rm max}^2 B \, {\rm Hz}$. A similar
requirement was also found by \citet{FalckMark00} for their RIAF-jet model. 
We find it difficult to fit the peak in the sub-mm excess feature at frequencies
$\gtapprox 100\,{\rm GHz}$.  It is likely that one or more of our simplifying
assumptions (e.g. keeping $\gamma_{\rm min}$ and $\gamma_{\rm max}$
constant throughout the jet)  may need to be relaxed in order to test more thoroughly
the jet interpretation of the sub-mm excess. However, the observational data are mostly upper 
limits in this band and there may be other contributions to this spectral component, 
such as warm dust emission \citep[e.g.][]{Becklin82}.

\begin{table}
\caption{Disk parameters used in the coupled disk-jet spectral model shown in Figure~\ref{datafig}. The
 dimensionless black hole spin is $a$, $\ej/\ea$ is the fraction of
 accretion power extracted by the magnetized jet, $\ea$ is the total
 accretion efficiency, $\ed$ is the disk radiative efficiency, $\mdot$
 is the mass accretion rate in ${\rm M_{\odot}\, {\rm yr}^{-1}}$ and
 $P_{\rm j}$ is the jet power in ${\rm erg \, s}^{-1}$. The black hole
 mass is fixed at $M = 3.7 \times 10^6 M_{\odot}$ and the total disk
 luminosity is fixed at $\Ld = 10^{-8} \LEdd \approx 5 \times
 10^{36} {\rm erg \, s}^{-1}$.}
\centerline{
\begin{tabular}{ccccccc}
\\
\hline\hline
\multicolumn{3}{c}{input parameters} && \multicolumn{3}{c}{inferred parameters}\\
\hline
a&$\ej/\ea$&$\ea$&&$\ed$&$\mdot$&$P_{\rm j}$ \\
\hline
$0.00$&$0.75$&$0.06$&&$0.02$&$5 \times 10^{-9}$&$1 \times 10^{37}$\\
$0.99$&$0.90$&$0.31$&&$0.03$&$3 \times 10^{-9}$&$5 \times 10^{37}$\\
\hline
\end{tabular}
}
\label{table1}
\end{table}

 The main difference between the high-spin and low-spin disk-jet models is that the $a=0.99$ model predicts a 
 larger jet kinetic power and a bluer disk spectrum compared to the $a=0$ model. 
 The larger jet kinetic power for the $a=0.99$ model arises because a spinning black 
 hole can extract more accretion power $P_{\rm a}$ than a non-spinning hole accreting 
 at the same rate (and note in Table~1 that the accretion rate for the $a=0.99$ model 
 is only slightly lower than that for the $a=0$ model, whereas the accretion efficiency is 
 more than a factor of $4$ higher). A bluer disk spectrum is predicted by the $a=0.99$ model 
 because the accretion disk extends all the way down to a last marginally stable orbit of 
 $\rms \approx 1.5 \rg$ (compared to $\rms \approx 6\rg$ for the $a=0$ model) and thus, the 
 peak temperature of the accretion disk is higher in the high-spin case than in the low-spin case. 
For an intrinsic disk 
luminosity $\Ld \approx 5 \times 10^{36} \, {\rm erg \, s}^{-1}$, we find 
$\theta_{\rm d} \gtapprox 87^\circ$ (i.e. the disk is almost edge-on) is required for the 
disk spectrum to fall below the observational upper limit in the near infra-red  
(see Fig.~\ref{datafig}). Importantly, this data point places an even stronger 
observational constraint on a standard accretion disk model, which predicts a 
higher $\Ld$ and therefore requires an even higher $\theta_{\rm d}$ for the same 
$\mdot$. This is because a standard disk model predicts that all the accretion 
power is radiated by the disk, whereas our jet-modified disk model takes into 
account magnetic extraction of accretion power by a jet and hence, predicts a 
dimmer and redder disk compared to that of a standard disk accreting at the same rate.

\section{Discussion}\label{Discussion}
\label{sec4}

We have demonstrated that the low radiative output from \SgrA is not
incompatible with a slowly accreting, geometrically thin cool disk. 
We have suggested that such a disk may form at small radii as a result 
of mass outflows and electron heating in the hot part of the accretion
flow. These effects divert binding energy from the ions, thus
reducing the scaleheight, allowing radiative cooling and Coulomb
collisions to operate on an inflow time. Numerical 
simulations of low-$\mdot$ accretion flows that include radiative
cooling are needed to test this prediction. 
We infer cold gas accretion rates of
$\mdot \simeq \mbox{a few} \times 10^{-9} {\rm M_{\odot}\, yr^{-1}}$. 
The disk radiative efficiency, $\ed \approx 0.02 - 0.03$, is lower 
than the total accretion efficiency $\ea$ because accretion 
power is extracted from the disk to form a jet. 

An important observational test of this
model is the predicted disk luminosity, which is difficult to
constrain observationally owing to the nearly edge-on orientation of the disk
\citep{Trippe07} and strong ($\sim 30\,$mag) interstellar extinction
\citep[see e.g.][]{Cuadra03}. 
We have shown here that an intrinsic disk luminosity $L_{\rm d}
\approx 5 \times 10^{36} {\rm erg \, s}^{-1}$ for \SgrA requires the
disk to be inclined at 
$\theta_{\rm d} \gtapprox 87^\circ$. 
Importantly, the constraint on $\theta_{\rm d}$ from this thin-disk model is not 
as stringent as that implied by a standard disk because a jet-modified disk is 
dimmer and redder as a result of efficient removal of accretion energy
to power the jet. 
The disk luminosity and inclination are constrained by so far
only a few flux measurements 
in the NIR. Clearly, more NIR observations of \SgrA in quiescence are
needed to place tighter constraints on the steady-state disk spectrum  
(see also the discussion below on jet X-ray emission). Applications of this model to 
other more suitable LLAGN may thus provide better model constraints.
So far, we have applied the model to M87 \citep{JK07} and have found
remarkably good agreement between the predicted and observed optical
spectra, as well as the predicted jet power and that inferred from
jet observations on kiloparsec scales. Furthermore, the model suggests that M87
may harbour a rapidly spinning black hole accreting at a rate of
$10^{-3} M_\odot \, {\rm yr}^{-1}$.

We have not attempted here to model the X-ray emission from \SgrA.
However, Compton processes could be included in the jet spectral model 
to calculate the  X-ray flux and further constrain the physical
parameters by comparing with X-ray observations. By considering
Compton scattering of the disk photons, in particular, we could in
principle be able to indirectly constrain the intrinsic disk
luminosity, $L_{\rm d}$, and jet-modified spectrum. We note, however,
that recent simultaneous NIR and X-ray observations of \SgrA
\citep{Hornstein07}, revealing a constant spectral index during
NIR flares and constant X-ray flux, may rule out inverse
Compton emission by jet electrons off disk photons in favour of
synchrotron self-comptonisation in the jet as the primary emission
mechanism for the X-rays. The constancy of X-ray emission during NIR
flares implies a separate source region for at least some of the NIR
emission. In our model, emission shortward of $2\mu$m is due primarily
to disk emission and some of the uncorrelated flaring
activity in this band  could arise from stochastic magnetic reconnection
events in the turbulent disk. On the other hand, NIR variability that is
correlated with variability in the submm and/or X-ray bands \citep[see
  e.g.][]{YZ06} can arise
in our model from prompt acceleration events in the jet that
temporarily raise the high-energy cutoff in the electron energy distribution.

\section{Conclusions}\label{Conclusions}

We have revisited theoretical constraints on thin-disk accretion
in \SgrA, showing that a geometrically-thick, hot, two-temperature
accretion flow cannot be sustained in the presence of outflows and
electron heating. We have revised observational constraints, taking into
consideration modified disk emission as a result of magnetic coupling
to a jet. The magnetic torque which drives the jet efficiently
extracts accretion power from the disk at small radii. This results in
an accretion disk that is dimmer than a standard
relativistic disk accreting at the same rate, 
so the constraints on disk luminosity and inclination are less stringent.
For \SgrA, we infer a mass accretion rate of
$\mdot \simeq \mbox{a few} \times 10^{-9} M_{\odot}\, {\rm yr}^{-1}$.  
We find that a disk 
luminosity of $\Ld \approx 5 \times 10^{36} \, {\rm erg \, s}^{-1} $ and an inclination 
angle $\gtapprox 87^\circ$ are compatible with existing observational constraints.  
However, more NIR observations of \SgrA in quiesence are needed to
provide tighter constraints on the presence of a cool accretion disk. 
Numerical simulations of radiative, low-$\mdot$ accretion flows 
are also warranted. 




\acknowledgments

The authors wish to thank the referee for comments and suggestions that helped to
improve the paper considerably.
EJDJ acknowledges support from a University of Sydney Postgraduate Award.
ZK acknowledges support from a University of Sydney Research Grant. 





\appendix
\section{Relativistic Disk Accretion}\label{A1}

The relativistic theory for steady-state, geometrically-thin disk
accretion onto a black hole was
formulated by \citet{b10} and \citet{b11}. In this formalism, the
radiative disk flux can be conveniently expressed as the Newtonian solution derived
by \citet{b9} multiplied by relativistic correction
factors. Specifically, for a
black hole with mass $M$ and dimensionless spin parameter $a$, accreting
at a rate $\mdot$, the radiative disk flux can be written as
\begin{equation}
F(x) = \frac{3}{8\pi} M^{-2} \mdot \, x^{-3} \, \fnt
\end{equation}
where  $r = M x$ is the Boyer-Lindquist radial coordinate and $\fnt$ is the
Novikov--Thorne relativistic correction factor, defined by
$$
\fnt = \frac{A(x)}{C(x)}\qquad , \qquad \rm{where} \qquad C(x) \equiv 1- 3x^{-1} + 2a x^{-3/2} \qquad \rm{and}
$$ 
\begin{eqnarray}
A(x) &\equiv& 1 - \frac{\yms}{y} -
\frac{3a}{2y}\ln\left(\frac{y}{\yms}\right) 
- \frac{3(y_1-a)^2}{yy_1(y_1 - y_2)(y_1 - y_3)}\ln\left(\frac{y -
  y_1}{\yms - y_1}\right) \nonumber \\
&-& \frac{3(y_2-a)^2}{yy_2(y_2 - y_1)(y_2 - y_3)}\ln\left(\frac{y -
  y_2}{\yms - y_2}\right) 
- \frac{3(y_3-a)^2}{yy_3(y_3 - y_1)(y_3 - y_2)}\ln\left(\frac{y - y_3}{\yms - y_3}\right)
\end{eqnarray}
with $y = x^{1/2}$ , $\yms = {\xms}^{1/2}$
and where
\begin{equation}
y_1 = 2\cos\left(\frac{1}{3}\cos^{-1}a - \frac{\pi}{3}\right) \qquad , \qquad
y_2 = 2\cos\left(\frac{1}{3}\cos^{-1}a + \frac{\pi}{3}\right) \qquad , \qquad
y_3 = -2\cos\left(\frac{1}{3}\cos^{-1}a \right)
\end{equation}
are the roots of the equation $y^3 - 3y + 2a = 0$.
The last marginally stable orbit in the Kerr metric is given by:
\begin{equation}
\xms = \left[3 + Z_2 - sign(a)\sqrt{(3 - Z_1)(3 + Z_1 + 2Z_2)}\right] 
\end{equation}
where
\begin{equation}
Z_1 = 1 + (1 - a^2)^{1/3}\left[(1 + a)^{1/3} + (1 - a)^{1/3}\right]
\qquad , \qquad
Z_2 = \sqrt{3a^2 + Z_1^2}
\end{equation}

\subsection{Correction factor for a nonzero torque at the inner boundary}

\citet{b12} calculated a correction term for the radiative flux from a
relativistic disk torqued at the inner boundary $\rms$.
In this case, the solution for the comoving disk flux can be generalized to
\begin{equation}
F(x) =  \frac{3}{8\pi} M^{-2} \mdot \, x^{-3} \, \left[ \, \fnt + \frphi  \,
  \right] 
\label{AKfluxeq}
\end{equation}
The correction factor for the nonzero torque at $\xms$ is
\begin{equation}
\frphi = \frac{C_{\rm ms}^{1/2}}{C(x)} \frac{2\pi \xms^{5/2} 
  T_{r\phi}(\rms) M \Omega_{\rm ms}}{x^{1/2} \mdot} 
\end{equation}
with $C_{\rm ms} = C (\xms)$, $\Omega_{\rm ms} = \Omega (\xms)$, where $\Omega  =  B^{-1} M^{-1} x^{-3/2}$ 
is the angular frequency of a circular orbit at radius $r$, with
$B(x) = 1 + a x^{-3/2}$
and where
\begin{equation}
T_{r\phi}(\rms) = \int_{-h}^{+h} t_{r\phi}(\rms) \, \mathrm{d}z
\end{equation}
is the vertically integrated radial-azimuthal magnetic stress at the
last marginally stable orbit \citep[see also][]{KB04}.

\subsection{Correction factor for a nonzero torque on the disk
  surface}\label{A2}

Here we derive a correction term for the radiative flux of a
relativistic disk with a nonzero magnetic torque on its surface. The
disk is torqued by open field lines that do work against the
surface. We identify this torque as being responsible for the
formation of a magnetized corona and/or jet and we can write the
generalized disk flux as
\begin{equation}
F (x) =  \frac{3}{8\pi} M^{-2} \mdot \, x^{-3} \, \left[ \, \fnt + \frphi - \fphiz
  \, \right] 
\label{e:Fx}
\end{equation}
where $\fphiz$ is the correction factor for the azimuthal-vertical
magnetic stress that gives rise to a nonzero torque on the disk
surface  \citep{KB04}.

We follow the procedure used by \citet{b11}, using slightly different
notation. We introduce the following parameters:
\begin{equation}
\xi = \frac{4\pi r F}{\mdot} \qquad , \qquad \varpi = \frac{2\pi r T_\phi^{\; r}}{\mdot}
\qquad , \qquad \Upsilon = \frac{4\pi r t_\phi^{+ \, z}}{\mdot}
\label{e:}
\end{equation}
where $F$ is the disk
flux measured in the fluid frame and $t_\phi^{+ \, z}$ is the magnetic
stress evaluated at the disk surface \citep{KB04}.

Conservation of angular momentum can be expressed as
\begin{equation}
\left[ L^\dagger  - \varpi \right]_{,r} \,
    = \, \xi L^\dagger \, +
    \, \Upsilon 
\label{e:angmom1}
\end{equation}
where $L^\dagger$ is the specific angular momentum of a circular orbit
at radius $r$. Energy conservation gives the relation
$
\varpi = B C^{1/2} \frac{\xi}{(-\Omega_{, \, r})} 
$
and substitution into (\ref{e:angmom1}) then gives
\begin{equation}
\left[ L^\dagger - B C^{1/2} \frac{\xi}{(-\Omega_{, \, r}) }  \right]_{,r} \,
    = \, \xi L^\dagger \, +
    \, \Upsilon 
\end{equation}
This is a first-order differential equation in $\xi$ with a general solution
\begin{equation}
\frac{ C }{B^2 (-\Omega_{, \, r})} \, \xi =  \int_{\rms}^r \frac{C^{1/2}}{B}
 \, \left( L^\dagger_{, \, r} - \Upsilon \right) \, {\rm d}r \, + \,
 \frac{C_{\rm ms}}{B_{\rm ms}^2 (-\Omega_{, \, r})_{\rm ms}} \xi_{\rm
 ms}
\label{e:soln}
\end{equation}
The last term on the right hand side of this equation takes into account
nonvanishing magnetic stresses at $\rms$. Energy conservation and the expression for 
$\varpi$ in (\ref{e:}) give
\begin{equation}
\xi_{\rm ms} = \frac{B_{\rm ms}}{C_{\rm ms}^{1/2}}  \frac{2 \pi \rms
T_\phi^{\; r} (\rms)}{\mdot} (\Omega_{, \,  r})_{\rm ms}
\end{equation}
This term is set to zero in the \citet{b11} treatment.

The term in (\ref{e:soln}) describing the effects of a nonzero suface
torque is the one involving $\Upsilon$. Using $(-\Omega)_{, \, r} =
\frac{3}{2} B^{-2} M^{-2} x^{-5/2}$ with $\xi$ and
$\Upsilon$ (see (\ref{e:})), the comoving energy flux removed from the disk by this magnetic torque is
\begin{equation}
F_{\phi z}^{\rm nzt} = \frac{3}{8\pi} M^{-3} \mdot x^{-7/2} C^{-1}
\int_{\rms}^r \frac{C^{1/2}}{B} \frac{4\pi r t_{\phi}^{+ \, z}}{\mdot}
\, {\rm d}r
\end{equation}
from which we deduce that the correction factor $\fphiz$
for a nonzero torque on the surface of a relativistic disk, as defined
in (\ref{e:Fx}), is
\begin{equation}
\fphiz = M \mdot^{-1} x^{-1/2} C^{-1} \int_{\xms}^x
\frac{C^{1/2}}{B} 4\pi x t_{\phi}^{+ \, z} \, {\rm d}x 
\end{equation}


------------------------------------------------------------------------

\clearpage

\end{document}